1

# Total Internal Reflection and the Related Effects[1]

(Translation of the Russian version appeared in the Proceedings of Tartu State University No. 42, p. 94-112, 1956)

*N. Kristoffel*
*Circle of theoretical physics. Supervisor docent Paul Kard*

## 1. Introduction

In 1909 Professor of Moscow University A.A.Eichenwald presented first the theory of total internal reflection basing on the electromagnetic theory of light [1]. Schafer and Gross following the method of Eichenwald have considered the passing of electromagnetic waves through a thin plate at the incident angle exceeding the limiting angle of the total reflection. They have reached accordance with the theory in the corresponding experiments [2]. Wiegrefe has studied the total reflection in the case of an arbitrary linear polarization of the incident light [3].

Recently principally novel experiments with the aim of proving the penetration of light into the second medium on total reflection have been performed by Goos and Hänchen [5,7]. In these investigations the shift of the totally reflected beam has been measured. The theory of this shift has first been developed in papers [6], [8].

The aim of the present contribution is to generalise the theory of total internal reflection for the case of elliptic polarization of the incident light.

Until now this general situation has been considered in few cases and the theories of shifts presented hitherto avoid it totally, being restricted to the linear polarization perpendicular or parallel to the plane of incidence.[2]

The total internal reflection is present under the condition

$$\sin \Theta > n = \frac{n_2}{n_1}, \quad (1.1)$$

where $\Theta$ is the angle of incidence, $n_1$ and $n_2$, the absolute refraction indices of the first and the second medium. The field in the second medium can be obtained from the usual Fresnel formulae[3] if one substitutes there

$$\sin \Theta'' = \frac{\sin \Theta}{n}, \quad (1.2)$$

$$\cos \Theta'' = -\frac{i}{n}\sqrt{\sin^2 \Theta - n^2}, \quad (1.3)$$

where $\Theta''$ is the complex refraction angle.

---

[1] Awarded a prize at the contest of Tartu State University (in April, 1954).
[2] Here and in what follows the direction of the linear polarization is taken being determined by the electric vector.
[3] The magnetic permeability ist taken equal to one.

## 2. The field in the optically sparse medium

Let us chose the interface between the media to be the $xy$ plane; the $z$ axis is directed downward, the $x$ axis, to the right, the $y$ axis, to the reader. Let the first medium be above and the second, below (see Fig. 1).

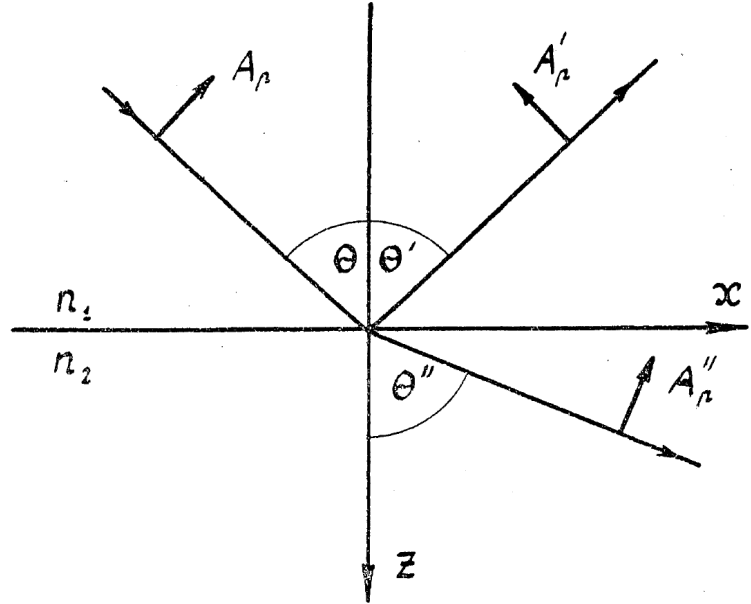

Fig.1.

The electric field of an infinite incident plane wave can be divided into two components – parallel and perpendicular to the plane of incidence. A distinct phase shift between them is present in the case of elliptic polarization. Therefore we write

$$\begin{cases} E_p = A_p e^{i(\omega t - \vec{k}\vec{r})} \\ E_y = A_y e^{i(\omega t - \vec{k}\vec{r} + \delta)} \end{cases}, \quad (2.1)$$

where $\omega$ is the frequency, $\vec{k}$ is the wave vector, $A_p$ and $A_y$ are the amplitudes of the components mentioned, and $\delta$, the phase shift. For the field in the medium of smaller index one can formally write analogous expressions

$$\begin{cases} E_p'' = A_p'' e^{i(\omega t - \vec{k}''\vec{r})} \\ E_y'' = A_y'' e^{i(\omega t - \vec{k}''\vec{r} + \delta)} \end{cases} \quad (2.1)$$

where $\vec{k}'' = nk$.

From the Fresnel formulae

$$\begin{cases} A_p^{"} = A_p \dfrac{2\cos\Theta\sin\Theta''}{\sin(\Theta+\Theta'')\cos(\Theta-\Theta'')} \\ A_y^{"} = A_y \dfrac{2\cos\Theta\sin\Theta''}{\sin(\Theta+\Theta'')} \end{cases} \qquad (2.3)$$

and taking account of (1.2) and (1.3), one finds

$$\begin{cases} A_p^{"} = A_p \dfrac{2n\cos\Theta}{\sqrt{(1-n^2)(\sin^2\Theta - n^2\cos^2\Theta)}} e^{i\psi} \\ A_y^{"} = A_y \dfrac{2\cos\Theta}{\sqrt{1-n^2}} e^{i\phi} \end{cases}, \qquad (2.4)$$

where

$$\begin{cases} \tan\psi = \dfrac{\sqrt{\sin^2\Theta - n^2}}{n^2\cos\Theta} \\ \tan\phi = \dfrac{\sqrt{\sin^2\Theta - n^2}}{\cos\Theta} = n^2\tan\psi. \end{cases} \qquad (2.5)$$

However, as further

$$\begin{cases} A_x^{"} = A_p^{"}\cos\Theta'' \\ A_z^{"} = -A_p^{"}\sin\Theta'', \end{cases}$$

then

$$\begin{cases} A_x^{"} = A_p \dfrac{2\cos\Theta\sqrt{\sin^2\Theta - n^2}}{\sqrt{(1-n^2)(\sin^2\Theta - n^2\cos^2\Theta)}} e^{i(\psi-\pi/2)} \\ A_y^{"} = A_y \dfrac{2\cos\Theta}{\sqrt{1-n^2}} e^{i\phi} \\ A_z^{"} = -A_p \dfrac{2\sin\Theta\cos\Theta}{\sqrt{(1-n^2)(\sin^2\Theta - n^2\cos^2\Theta)}} e^{i\phi}. \end{cases} \qquad (2.6)$$

The components of the magnetic vector in the second medium will be calculated from the second Maxwell equation. Introducting the designations

$$\kappa = k\sqrt{\sin^2\Theta - n} \qquad (2.7)$$

$$\tau = \omega t - k\sin\Theta x, \qquad (2.8)$$

and going to real quantities, the electromagnetic field in the second medium is obtained in the form

$$E_x^{"} = 2\cos\Theta\sin\psi A_p e^{-\kappa z}\sin(\tau+\psi)$$
$$E_y^{"} = 2\cos\phi A_y e^{-\kappa z}\cos(\tau+\delta+\phi)$$
$$E_z^{"} = -\dfrac{2}{n^2}\sin\Theta\cos\psi A_p e^{-\kappa z}\cos(\tau+\psi)$$

$$H_x^{''} = -2n_1 \cos\Theta \sin\varphi A_y e^{-\kappa z} \sin(\tau + \delta + \phi)$$
$$H_y^{''} = 2n_1 \cos\psi A_p e^{-\kappa z} \cos(\tau + \psi)$$
$$H_z^{''} = 2n_1 \sin\Theta \cos\phi A_y e^{-\kappa z} \cos(\tau + \delta + \phi).$$

(2.9)

In addition to the periodic alteration in time and along the $x$ axis, the field in the second medium decreases exponentially with moving off from the separation plane. The exponential factor $e^{-\kappa z}$ can be presented in the form $\exp\left(-\frac{2\pi}{\lambda_1}z\sqrt{\sin^2\Theta - n^2}\right)$ and if $\sqrt{\sin^2\Theta - n^2}$ is not too small, a remarkable weakening of the field in the second medium occurs. Already at the $z$-s that are comparable to the wavelengths of light, the field in the second medium decreases quite rapidly.

Only for the angles of incidence close to the limiting one, the electromagnetic field in the medium of smaller index has noticeable values at relatively large depths.

The calculation of the field of the reflected wave shows that in the reflected wave there appears a phase shift with respect to the incident wave, which equals $2\psi$ for the $\parallel$-component and $2\varphi$ for the $\perp$-component.[4]

### 3. Energy flux curves in the optically sparse medium

To investigate the movement of the energy in the second medium we use the Umov-Poynting vector:

$$\vec{S}'' = \frac{c}{4\pi}(\vec{E}' \times \vec{H}'').$$

(3.1)

Substituting here (2.9), we arrive at

$$S_x^{''} = \frac{cn_1 \sin\Theta \cos^2\Theta}{\pi(1-n^2)} e^{-2\kappa z}\left[A_y^2 \cos^2(\tau + \delta + \varphi) + \frac{n^2}{\sin^2\Theta - n^2 \cos^2\Theta} A_p^2 \cos^2(\tau + \psi)\right]$$

$$S_y^{''} = \frac{cn_1 A_p A_y}{\pi\sqrt{1-n^2}} \sin\Theta \cos^2\Theta \sin\psi \sin(\delta + \varphi - \psi) e^{-2\kappa z}$$

(3.2)

$$S_z^{''} = \frac{cn_1 \cos^2\Theta \sqrt{\sin^2\Theta - n^2}}{2\pi(1-n^2)} e^{-2\kappa z}\left[A_y^2 \sin 2(\tau + \delta + \varphi) + \frac{n^2}{\sin^2\Theta - n^2 \cos^2\Theta} A_p^2 \sin 2(\tau + \psi)\right]$$

We stress that the y-component of the Umov-Poynting vector is nonzero and does not depend on time. It means that the movement of the energy in the direction perpendicular to the plane of incidence is stationary.[5]

Averaging the z-component of the Umov-Poynting vector in time, we obtain zero

---

[4] $\parallel$ - the component in the plane of incidence; $\perp$ - the component perpendicular to the plane of incidence.

[5] We mentione M.Born's mistake supposing in [4] that $S_y^{''} = 0$ (p. 62).

$$\vec{S}''_y = 0 \qquad (3.3)$$

This means that the amount of energy flowing through the interface into the second medium equals to the one returning from there. In other words, the total internal reflection is really „total": all the energy entered into the second medium, returns to the first medium with no losses.

Especially significant is the circumstance that the light intensity in the second medium depends on the polarization of the incident light (see (3.2)).

Eichenwald has stated the opposite. For a linearly polarized light (in the plane of incidence) Eichenwald finds the following components of the Umov-Poynting vector (using the designations of Eichenwald):

$$\begin{cases} f_x = A^2 \dfrac{\varepsilon a}{8\pi k^2} e^{-\frac{4\pi x}{\tau}z} \left[1 - \cos\dfrac{4\pi}{\tau}(t-ax)\right] \\ f_z = -A^2 \dfrac{\varepsilon}{8\pi k} e^{-\frac{4\pi x}{\tau}z} \sin\dfrac{4\pi}{\tau}(t-ax) \end{cases}$$

For a light polarized perpendicular to the plane of incidence Eichenwald presents no formulae in his paper, however these one can be easily calculated:

$$\begin{cases} f_x = B^2 \dfrac{\varepsilon a}{8\pi k^2}(\sin^2\varphi - n^2) e^{-\frac{4\pi x}{\tau}z} \left[1 - \cos\dfrac{4\pi}{\tau}(t-ax)\right] \\ f_z = -B^2 \dfrac{\varepsilon}{8\pi k}(\sin^2\varphi - n^2) e^{-\frac{4\pi x}{\tau}z} \sin\dfrac{4\pi}{\tau}(t-ax) \end{cases}.$$

These formulae differ from the preceeding ones firstly, by the factor $\sin^2\varphi - n^2$ and, secondly, instead of $B$ there stands the amplitude $A$. Eichenwald states that at $A = B$ both formulae must coincide, but it is not so, which is one side of Eichenwald's mistake. The other one consists in the fact that $A = B$ does not mean the equality of the amplitudes in the incident wave, which is what Eichenwald apparently supposes.

Now we find the energy flux curves in the second medium. These are the curves to which the Umov-Poynting vector is tangent at all time moments.

For the projection of the flux curves on the $xz$ plane we obtain the following differential equation:

$$\frac{dz}{dx} = \frac{S''_z}{S''_x} = \frac{\sqrt{\sin^2\Theta - n^2}}{2\sin\Theta} \times \frac{A_y^2 \sin 2(\tau + \delta + \phi) + \dfrac{n^2 A_p^2}{\sin^2\Theta - n^2\cos^2\Theta}\sin 2(\tau + \psi)}{A_y^2 \cos^2(\tau + \delta + \phi) + \dfrac{n^2 A_p^2}{\sin^2\Theta - n^2\cos^2\Theta}\cos^2(\tau + \psi)}. \qquad (3.4)$$

Analogously for the projections of the flux curves on the $xy$ plane

$$\frac{dy}{dx} = \frac{S_y^{"}}{S_x^{"}} = \frac{A_p A_y \sqrt{1-n^2}\, \sin\psi \sin(\delta+\phi-\psi)}{A_y^2 \cos^2(\tau+\delta+\phi) + \dfrac{n^2 A_p^2 \cos^2(\tau+\psi)}{\sin^2\Theta - n^2 \cos^2\Theta}}. \tag{3.5}$$

Integrating equations (3.4) and (3.5) we obtain

$$\zeta = \ln(C' + \cos 2\xi) + const$$
$$\eta = -\arctan\left(\sqrt{\frac{C'-1}{C'+1}}\tan\xi\right) + const, \tag{3.6}$$

where

$$\xi = k\sin\Theta x$$
$$\eta = \frac{y}{A} \tag{3.7}$$
$$\zeta = \frac{z}{B}.$$

$A$ and $B$ are constants not depending on coordinates, the constant $C'$ is determined as follows

$$C' = \frac{2\left(A_y^2 + \dfrac{n^2 A_p^2}{\sin^2\Theta - n^2\cos^2\Theta}\right)}{\sqrt{A_y^4 + \dfrac{n^4 A_p^4}{(\sin^2\Theta - n^2\cos^2\Theta)^2} - \dfrac{2A_p^2 A_y^2 n^2 \cos^2\left(\delta+\phi-\psi+\dfrac{\pi}{2}\right)}{\sin^2\Theta - n^2\cos^2\Theta}}}. \tag{3.8}$$

Expressions (3.6) hold for one time moment. The flux curves calculated from (3.6) are shown in Figs. 2 and 3.

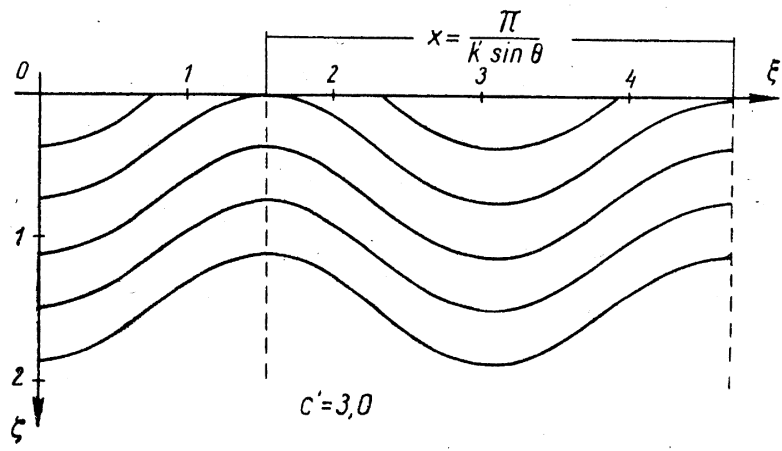

Fig.2.

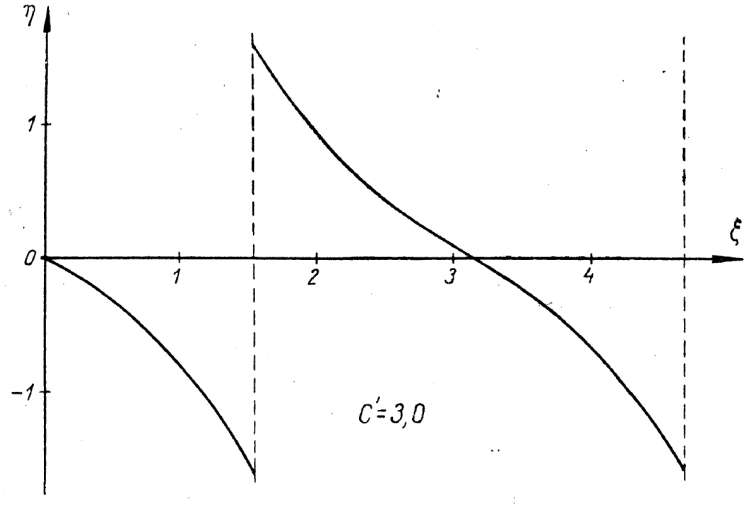

Fig.3.

As can be seen, the projections of the flux curves on the $xz$ plane are of the wave form. With the increasing $C'$ they become ever more flat. If at the given moment the energy at a point $x = 0$ flaws into the second medium, then at the same moment at the distance $x = \dfrac{\pi}{k \sin \Theta}$ the energy is coming out from the second medium. It means that on total reflection the light beam undergoes a shift along $x$ axis. However, the real shift of the beam is not equal to the expression mentioned because the image of the flux curves depends also on time.

In the direction of the $z$ axis, the energy oscillations have periods twice smaller than those along the $x$ axis, i.e. during one oscillation in the incident light the energy has time to enter twice the second medium and to come out from there.

Of special interest is the perpendicular shift of the beam, the so-called $y$-shift, the existcne of which follows from the projections of the flux lines onto the interface ($\xi\eta$). If at the given moment the energy flows into the second medium at a point $x$, then it comes out from there at another time moment, but already not on the $x$ axis. Thus the plane of incidence and that of reflection, though being parallel, do not coincide. The $y$-shift was first paid attention to by Kard (private communication).

Of some interest is the case, where $C' = \infty$. According to (3.8) the condicitions for this are

$$A_y = \pm \frac{nA_p}{\sqrt{\sin^2 \Theta - n^2 \cos^2 \Theta}}, \quad \delta + \varphi - \psi = \pm \frac{\pi}{2}. \qquad (3.9)$$

From (2.5) the corresponding polarization of the incident light is found

$$\tan \delta = -\frac{\sin^2 \Theta}{\cos \Theta \sqrt{\sin^2 \Theta - n^2}}. \tag{3.10}$$

Two solutions of (3.10) for $\delta$ correspond to two rotation possibilities on the polarization ellipse. For the projections of flux lines one now obtains

$$\begin{aligned} \eta &= -\xi + const \\ \zeta &= const \end{aligned}. \tag{3.11}$$

The flux curves degenerate into straight lines. In the plane, the flux curves are straight lines, parallel to the $\xi$ axis and in $\xi\eta$ plane straight lines with the curvature angle -45°.

Judging by the flux curves, the energy flows in the medium of smaller index parallel to the interface and leaves it only in the infinity. The problem consists in the circumstance that this result belongs only to an infinite incident wave while in experiments we always have restricted light beams. The flux lines will not be straight and the distortions appear which show the places of in- and outcoming of the energy.

The dependence of flux curves on time consists in the movement of the whole picture in the positive $x$ direction with the constant velocity $u$

$$u = \frac{\omega}{k \sin \Theta} = \frac{c}{n_1 \sin \Theta}. \tag{3.12}$$

The investigation of the energy movement in the second medium is possible by using the hodograph of the Umov-Poynting vector. The corresponding calculations show that the hodograph is an ellipse lying in the plane parallel to the plane of incidence.

**4. Papers on beam shifts on total reflection**

Artmann in [6] was the first to provide the theory of shifts for linearly polarized light. He supposed the beam to be finite, as in his opinion the shift is caused especially by the finiteness of the incident beam. For the shift $D$ perpendicular to the beam in the plane of incidence Artmann obtained the expression

$$D = \frac{\lambda_1}{2\pi} \frac{d\chi}{d\Theta}, \tag{4.1}$$

where $\chi$ is the shift on total reflection (with our designations $\chi_\perp = 2\varphi$, $\chi_\parallel = 2\psi$)[6]. Taking into account of (2.5),

$$D_\perp = \frac{\lambda_1}{\pi} \frac{n}{\sqrt{\sin^2 \Theta - n^2}} \tag{4.2}$$

---

[6] Artmann writes formula (4.1) with a minus sign, which is connected with another choice of complex representation.

$$D_{\parallel} = \frac{\lambda_1}{\pi} \frac{1}{n\sqrt{\sin^2\Theta - n^2}}. \qquad (4.3)$$

The indices $\perp$ and $\parallel$ designate the polarization of the incident light perpendicular to the plane of incidence and in this plane.

The shortcoming of the Artmann formulae is that it is justified only for the angles of incidence close to the limiting one (therefore in (4.2) and (4.3) also $\sin\Theta \approx n$ is used).

Fragstein [8] supposed that the amplitude in the incident beam decreases monotonically at its edges and he obtained the same result. From (4.2) and (4.3) it can be seen that the shift depends on the polarization of the incident light. As in the first experiments [5] this dependence was not found, the necessity of new more precise experiments become clear. A repeated work [7] gave the results that were in good agreement with the theory.

H.Wolter has observed a shift of the minimum of interference for two plane waves incident on the interface with angles $\Theta$ and $\Theta - \Delta$, where $\Delta$ is a small angle [9]. His experiment was in good agreement with the theory and the dependence of the shifts on polarization was confirmed once more.

A more simple deduction of formula (4.1) has been performed in [11] by L.M.Brekhovskikh but the general prepositions of Brekhovskikh's method coincide with those made by Artmann.

We are interested in the problem of shifts on the elliptic polarization of the incident light, about which there are no data in literature yet.

According to the Artmann-Brekhovskikh method that shift of the beam depends on the phase shift. However, as for the $\parallel -$ and $\perp -$ components of the elliptically polarized light the phase shifts are different, the beam shifts for these polarizations are to be calculated separately. As $n < 1$, according to (4.2) and (4.3) we have $D_{\parallel} > D_{\perp}$, i.e. the shifts of the $\parallel -$ and $\perp -$ components on the total reflection of the elliptically polarized light are not equal. If the beam is narrow, the reflected beam must correspondingly split into two components.

In [7], there is a short note on an experimental observation of this splitting. However, as will be seen below the theoretical convincingness of this effect is questionable and, therefore, new experiments are needed to prove it.

A drawback of the Artmann-Brekhovskikh method is the fact that the field in the optically sparse medium has not been considered there, whereas the total internal reflection is a process taking place in both media and these must be considered together.

Concerning the supposed $y$-shift, the Artmann-Brekhovskikh method is unable to predict it, as the existence of the $y$-shift is deduced from the picture of the energy movement in the second medium, which is absent in the Artmann-Brekhovskikh method. The $y$-shift is a characteristic effect of the general linear and elliptic polarization (ch. 5), while the Artmann-Brekhovskikh method is applicable only to the $\parallel -$ and $\perp -$ components separately.

## 5. Calculation of shifts according to the method of P.Kard

In the papers concerning total reflection, until now no account has been of the significant circumstance that the beam shifts on total reflection must depend on the intensity of the energy flux in the second medium. Indeed, the higher the intensity of the light is, the deeper the light penetrates into the second medium and the larger should be the shift of the beam returning into the first medium. The field in the second medium decreases with the moving off from the interface according to the factor $e^{-\kappa z}$, where $\kappa = \frac{2\pi}{\lambda_1}\sqrt{\sin^2\Theta - n^2}$. The quantity $1/\kappa$, independent of polarization is usually considered as the depth of the light penetration into the second medium. However, already the earlier investigations by Quincke showed that the depth of the light penetration into the second medium depends on polarization.

A.A.Korobko-Stefanov in [10] considers the reason for this contradiction to be in the circumstance that the theory giving the factor $e^{-\kappa z}$ holds only for an unrestricted evanescent wave, but the experiments are performed with finite beams.

However, the problem can be elucidated more easily. In Fig.4, the dependence of the relative light intensity in the second medium on the distance to the interface for the light polarized in the plane of incidence and perpendicular to it is shown. At the $z = \frac{1}{2\kappa}$ the intensity is diminished by $e$ times but for different polarizations the intensities are different (also for $z = \frac{1}{\kappa}$). Consequently, the factor $e^{-\kappa z}$ does not give unique the depth of the light penetration into the second medium and the definition of the latter needs refinement. As the depth of penetration into the second medium $h$ we understand a distance from the interface at which the relative intensity of the light equals $1/e$. The relative intensity is the ratio of the average value of the $x$ component of the Umov-Poynting vector in the second medium to the absolute value of the Umov-Poynting vector in the incident light. As the intensity of the light in the second medium depends on polarization, the penetration depth defined in such a manner also depends on the polarization.

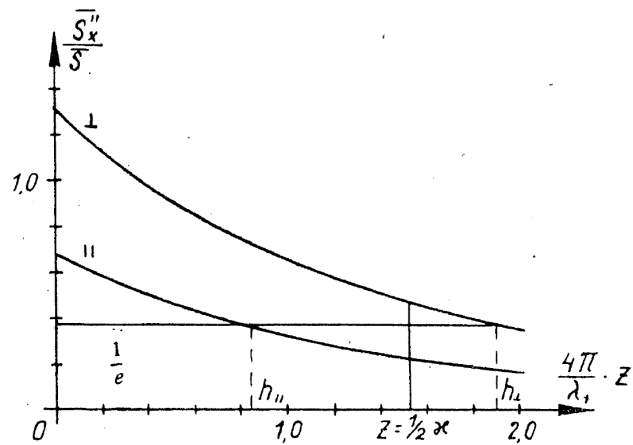

Figure 4

Let us find the depth of light penetration into the second medium for an elliptically polarized incident light. From (3.2) one obtains

$$\bar{S}_x'' = \frac{cn_1 \sin\Theta \cos^2\Theta}{2\pi(1-n^2)} \left[ A_y^2 + \frac{n^2 A_p^2}{\sin^2\Theta - n^2\cos^2\Theta} \right] e^{-2\kappa z}. \tag{5.1}$$

The relative intensity of the light in the second medium is found by dividing (5.1) by $\bar{S} = \frac{cn_1}{8\pi}(A_y^2 + A_p^2)$. According to our definition at $z = h$, $\bar{S}_x''/\bar{S} = e^{-1}$. One obtains

$$h = \frac{\lambda_1}{4\pi\sqrt{\sin^2\Theta - n^2}} \left[ \ln\left( \frac{4\sin\Theta\cos^2\Theta}{1-n^2} \times \frac{A_y^2 + \frac{n^2 A_p^2}{\sin^2\Theta - n^2\cos^2\Theta}}{A_y^2 + A_p^2} \right) + 1 \right]. \tag{5.2}$$

The penetration depth is proportional to the wavelength of the incident light. From (5.2), taking $A_p = 0$ or $A_y = 0$, the penetration depths for a correspondingly linearly polarized light are found; $h_\perp$ and $h_\parallel$ are shown in Fig. 4 for the case $\Theta = 60^\circ$, $n = \frac{1}{\sqrt{3}}$. If $h_\perp = h_\parallel$, then

$$\Theta = \Theta_0 = \arccos\sqrt{\frac{1-n^2}{1+n^2}}. \tag{5.3}$$

The value of $\Theta_0$ calculated from this expression is in good agreement with the value of $\Theta_0$ found by Quincke in his measurements.[7]

The account of the intensity of the energy flux in the second medium lies in the basis of an original method of calculating the shifts, proposed by P. Kard (private communication). Kard's method is immediately applicable to an elliptically polarized incident light.

Let take in the second medium a section of an unit width along the $y$ axis, that is perpendicular to the $x$ axis. The amount of energy flowing through this section in the unit of time is given by

$$w_x'' = \int_0^\infty \bar{S}_x'' dz. \tag{5.4}$$

Taking account of (5.1), we find

$$w_x'' = \frac{\lambda_1 n_1 c \sin\Theta \cos^2\Theta}{8\pi^2(1-n^2)\sqrt{\sin^2\Theta - n^2}} \left[ A_y^2 + \frac{n^2 A_p^2}{\sin^2\Theta - n^2\cos^2\Theta} \right]. \tag{5.5}$$

Further we suppose that the whole incident energy, before returning into the first medium in the form of reflected energy, passes through the second medium. Then the shift can be found as such a width of the incident light beam $D$, which bears the same amount of energy $w_x''$. Indeed, the amount of energy contained in the beam of such width flows through the section of the second medium and after reflection this beam must occur on the other side of the section. As can be seen (see Fig. 5), the shift of each ray in the beam owing to this just equals to this width.

---

[7] See [10], where the results of Quincke are given.

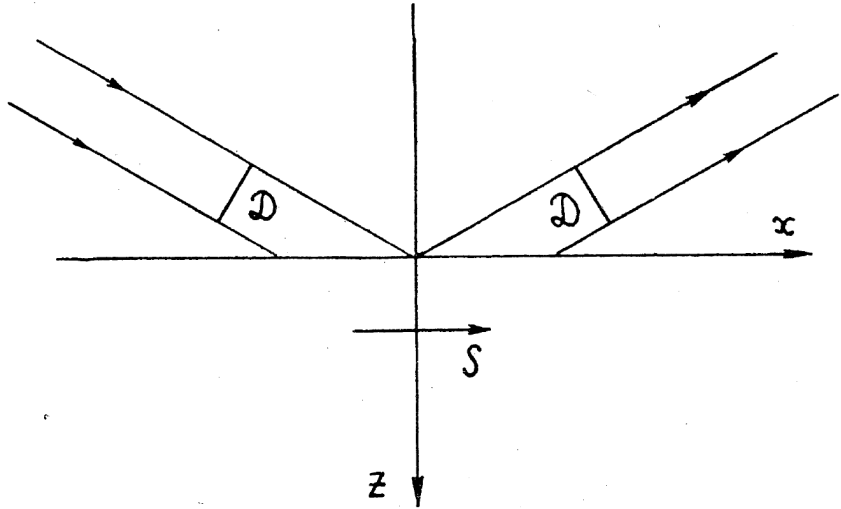

Figure 5

The shift being looked for is found by dividing (5.5) by $\overline{S}$

$$D_x = \frac{\lambda_1}{\pi} \cdot \frac{\sin\Theta \cos^2\Theta}{(1-n^2)\sqrt{\sin^2\Theta - n^2}} \cdot \frac{A_y^2 + \dfrac{n^2 A_p^2}{\sin^2\Theta - n^2 \cos^2\Theta}}{A_y^2 + A_p^2} \qquad (5.6)$$

When deriving this formula we do not suppose that the angle of incidence must be close to the limiting one and therefore the expression (5.6) must be applicable to any angle of incidence in the total reflection region.

In a such manner the beam shifts on total reflection can be calculated even from the theory of the infinite evanescent wave. If the incident beam is sufficiently broad, as our calculations have shown, the account of diffraction introduces but a small correction of the order of $\lambda_1^2/a^2$, where $a$ is the slit width. Thus, the account of diffraction introduces nothing new into Kard's method and therefore formula (5.6) may immediately be applied to experimental data.

As according to (5.6) (see also Fig.6, where $\Delta_x = \dfrac{D_x}{\cos\Theta}$), the shift is more remarkable near the limiting angle. We substitute $\sin\Theta \approx n$ into (5.6). Then

$$D_x = \frac{\lambda_1}{\pi} \cdot \frac{n}{\sqrt{\sin^2\Theta - n^2}} \cdot \frac{A_y^2 + \dfrac{A_p^2}{n^2}}{A_y^2 + A_p^2}. \qquad (5.7)$$

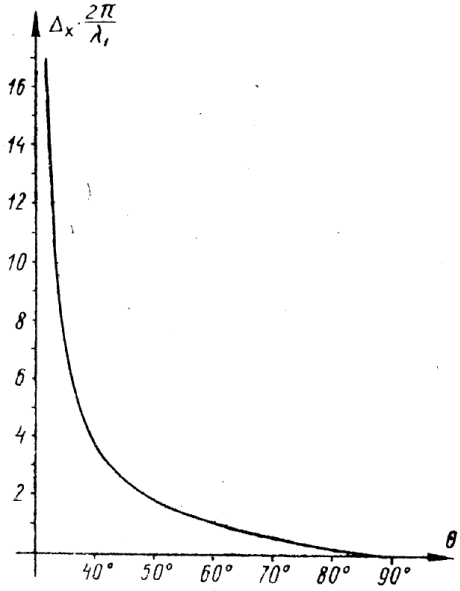

Figure 6

From here, taking $A_y = 0$ or $A_p = 0$, we arrive at Artmann-Brekhovskikh's formulae. Consequently formula (5.6) satisfies the experimental data (in the case of linear polarization) as well as does the Artmann-Brekhovskikh's.

Therefore a problem arises which of these contradicting rsults is the correct one. The answer to it can supposedly be obtained from new experiments on shift with an elliptically polarized light with the aim to confirm or to disprove the splitting.

Also, experiments of other type are possible, namely, on the predicted $y$-shift. As we saw in ch. 3, the appearance of such a shift is deduced from the presence of the corresponding energy flux in the second medium. The application of Kard's method in this case gives

$$D_y = \frac{\lambda_1}{\pi} \cdot \frac{2\sin\Theta\cos\Theta\sin\psi\sin(\delta+\phi-\psi)}{\sqrt{(1-n^2)\sin^2\Theta-n^2}} \cdot \frac{A_p A_y}{A_y^2 + A_p^2}, \qquad (5.8)$$

or

$$D_y = \frac{\lambda_1}{\pi} \cdot \frac{2\sin\Theta\cos\Theta}{(1-n^2)(\sin^2\Theta-n^2\cos^2\Theta)} \cdot \frac{\alpha}{1+\alpha^2}(\sin^2\Theta\sin\delta - \cos\Theta\cos\delta\sqrt{\sin^2\Theta-n^2}), \quad (5.9)$$

where

$$\alpha = \frac{A_p}{A_y}.$$

At $A_y = 0$ or $A_p = 0$ we have $D_y = 0$. But for a general linear or elliptic polarization $D_y \neq 0$. For fixed $\Theta$, the maximum of $D_y$ is at $\alpha = 1$, $\tan\delta = -\dfrac{\sin^2\Theta}{\cos\Theta\sqrt{\sin^2\Theta-n^2}}$. Then

$$D_{y\max} = \frac{\lambda_1}{\pi} \cdot \frac{\sin\Theta\cos\Theta}{(1-n^2)\sqrt{\sin^2\Theta - n^2\cos^2\Theta}} \ . \tag{5.10}$$

The largest value of (5.9) is obtained at $\sin\Theta \approx n$. Then

$$D_{y\max} = \frac{\lambda_1}{\pi} \cdot \frac{1}{n\sqrt{1-n^2}} \ . \tag{5.11}$$

## 6. On the possibility of an experimental discovery of $y$-shift

As the $y$-shift is remarkably smaller than the $x$-shift, it would be more difficult to detect it experimentally. We consider here two possible experiments for discovering the $y$-shift.

Like in the experiments of $x$-shift, the reflection must be multiple. And as $D_y$ depends on the phase shift between the $\perp$- and $\parallel$-components and this shift diminishes after each reflection by $\chi = 2\psi - 2\varphi$, the calculation of the total shift is much more complicated. According to (2.5)

$$\tan\chi = \frac{\cos\sqrt{\sin^2\Theta - n^2}}{\sin^2\Theta} \ . \tag{6.1}$$

In the expression for $D_y$ there is the term $\sin(\delta + \varphi - \psi)$ to which $D_y$ is proportional. Further, $\sin(\delta + \varphi - \psi) = \sin(\delta - \chi/2)$. At a single reflection the argument of this sine changes by $\chi$. In view of the monotonic decrease of $\delta$ with each reflection, $D_y$ changes the sign after a sufficient number of reflections. Therefore a significant difficulty arises, as it may happen that the shifts cancel each other and we do not reach the desired increase of this shift.

One must prepare such reflection conditions that at the moment when $D_y$ changes its sign the direction of reflection is also changed. Then the shifts are added. This is the basic idea of the first experiment proposed. The corresponding conditions can be reached by a special constructed prism.

Let us have conditions where after a three-fold reflection $D_y$ changes its sign. Let us take a four-faced prism with a rectangular basis with the edges in proportion 1:2. Then directing the beam into the prism under the incident angle of 45° at a ¼ distance of the length of the longer edge from the side-edge, the beam is reflected in the prism for 6 times successively while initially 3 times to the left and then 3 times to the right (following the direction of the beam), as shown in Fig. 7. The beam must enter the prism under a small angle in the vertical plane so that on multiple successive reflections it would climb higher and higher in the prism from where after a sufficient number of reflections it is let out onto a screen (photoplate). To let the beam in and out from the prism small isosceles prism glued onto the large prism can be used.

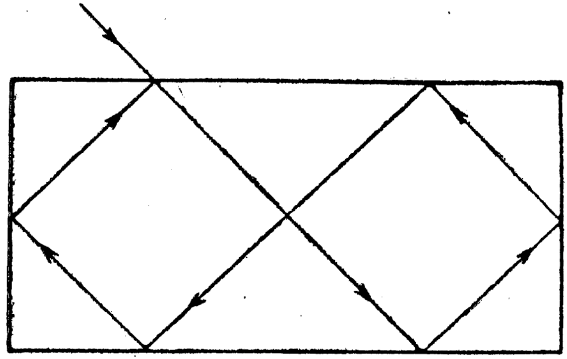

Figure 7

It is clear that under such conditions the following requirements must be satisfied: $\delta = 180°$, $\chi = 60°$ and $\chi/2 = 30°$. The way how the phase shifts change in case of a sixfold reflection is shown in Fig.8. The dotted lines denote $\sin(\delta - \chi/2)$.

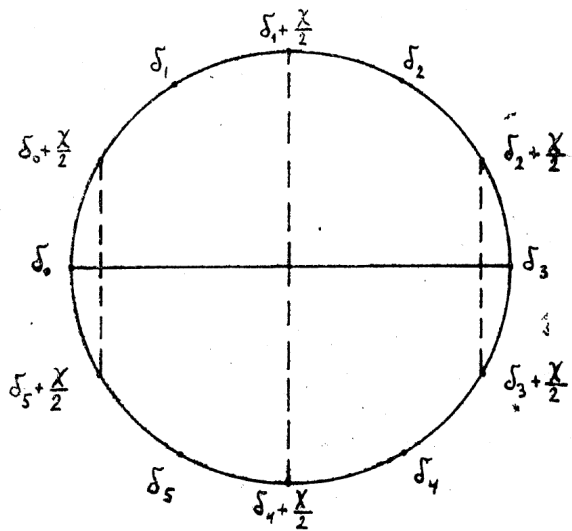

Figure 8

Beginning with the point $\delta_3$, $D_y$ changes sign, however, as at the same time the direction of reflection is changed, the shifts add. No such conditions can be obtained with any free refraction index. The suitable $n$ is found from the expression for $\tan \chi/2$, which is now convected into an equation for $n$.

Substituting $\Theta = 45°$, $\chi/2 = 30°$ into (6.1), one finds $n = \dfrac{1}{\sqrt{3}}$. Entering the beam into the prism, we can see its trace on the screen. When the polarization in the incident light changes due to the change of the $D_y$ sign, the trace of the beam on the screen must be displaced, which would be the experimental detection of the predicted transversal $y$-shift.

The experiment described can also be performed under other conditions. For instance, if $D_y$ changes its sign after 5 reflections, one must use a fourfaced prism, the basing edges of which are in the proportion 2:3. To $\delta = 180^o$ corresponds then $\chi/2 = 18^o$ and $n = \frac{1}{\sqrt{3}}$. The realization of such an interesting experiment involves difficulties due to the complexity of preparing suitable prisms.

Also, the following experiment is possible. The light beam is incident on a prism with a cancave spherical plane, which is placed into a liquid with the refraction index somewhat exceeding the refraction index of the prism, so that on the spherical plane there occurs total internal reflection (see Fig. 9).

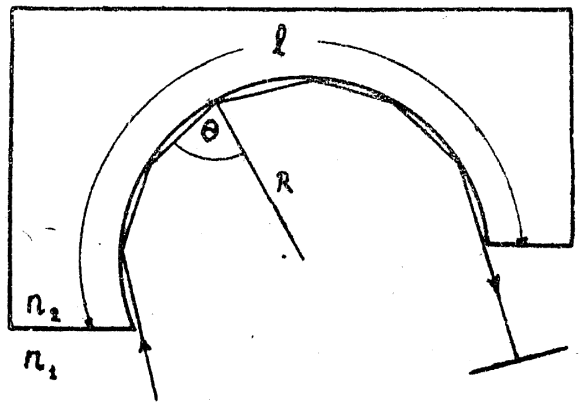

Figure 9

As in such conditions $n$ is close to one, the angle of incident must be close to 90° and before falling onto the screen (photoplate) the beam will reflect many times on the spherical plane. The size of the prism must be chosen so that $m \cdot \chi < \delta$, where $m$ is the number of reflections. Then, in case of all reflections we obtain positive $D_y$, which will be added. The light, falling to the screen, forms a kind of diffraction pattern there. When the polarization direction in the incident light changes, the diffraction pattern must displace in the vertical plane, with which the $y$-shift would be detected.

For a prism prepared from crown glass ($n_2 = 1.550$) in nitrobenzene $C_6H_5NO_2$ ($n_1 = 1.553$), $n = 0.999$. To the angle of incidence $\Theta = 88^o$ ($\sin \Theta = 0.9994$) corresponds $\chi = 6'$ and if in the incident light $\delta = 180^o$, we can use 1800 reflections, as in these conditions $D_y$ does not change its sign.

However, such a large number of reflections cannot be used as the spherical plane cannot space them (certainly, one may arrive at rising the beam as in the preceeding experiment to increase the number of reflections). If one restricts ourself to 80 reflections, then taking the radius of the spherical plane curvature as 5 cm, the length of the arc used ($l$ in fig.9) must be about 17 cm. Under such conditions according to formula (5.11), the observable displacement of the diffraction patterns should be approximately 0.6 mm ($\lambda_1 = 6 \cdot 10^{-5}$ cm), which is sufficiently well-observable quantity.

To perform the experiment one should use a light of a possibly large wavelength. Certainly, various modifications of this experiment are possible with the corresponding choice of prisms and liquids.

The latter experiment seems to be the easiest for the experimental detection of the $y$-shift.